\documentclass[prd,twocolumn,nofootinbib]{revtex4}
\usepackage{graphicx}
\usepackage{latexsym}

\begin{document}

\title{Constraining DGP Gravity from Observational Data}

\author{Jun-Qing Xia}
\email{xia@sissa.it}

\affiliation{Scuola Internazionale Superiore di Studi Avanzati, Via
Beirut 2-4, I-34014 Trieste, Italy}


\begin{abstract}

The accelerating expansion of our universe at present could be
driven by an unknown energy component (Dark Energy) or a
modification of general relativity (Modified Gravity). In this note
we revisit the constraints on a phenomenological model which
interpolates between the pure $\Lambda$CDM model and the
Dvali-Gabadadze-Porrati (DGP) braneworld model with an additional
parameter $\alpha$. Combining the cosmic microwave background (CMB),
baryon acoustic oscillations (BAO) and type Ia supernovae (SNIa), as
well as some high-redshift observations, such as the gamma-ray
bursts (GRB) and the measurements of linear growth factors (LGF), we
obtain the tight constraint on the parameter $\alpha=0.254\pm0.153$
($68\%$ C.L.), which implies that the flat DGP model is incompatible
with the current observations, while the pure $\Lambda$CDM model
still fits the data very well. Finally, we simulate the future
measurements with higher precisions and find that the constraint on
$\alpha$ can be improved by a factor two, when compared to the
present constraints.

\end{abstract}


\maketitle


\section{Introduction}

Current cosmological observations, such as the CMB measurements of
temperature anisotropies and polarization at high redshift
$z\sim1090$ \cite{Komatsu:2008hk} and the redshift-distance
measurements of SNIa at $z<2$ \cite{Kowalski:2008ez}, have
demonstrated that the universe is now undergoing an accelerated
phase of expansion. The simplest explanation is that this behavior
is driven by the cosmological constant or the dynamical dark energy
models, which suffers from the severe coincidence and fine-tuning
problems \cite{CCproblem}. On the other hand, this observed
late-time acceleration of the expansion on the large scales could
also be caused by some modifications of general relativity, such as
the scalar-tensor \cite{ST} and $f(R)$ theories \cite{fR}, and
gravitational slip \cite{slip}.

One of the well-known examples is the DGP braneworld model
\cite{DGP}, in which the gravity leaks off the four dimensional
brane into the five dimensional space-time. On small scales gravity
is bound to the four dimensional brane and the general relativity is
recovered to a good approximation. In the framework of flat DGP
model, the Friedmann equation will be modified as
\cite{DGPCosmology}:
\begin{equation}
H^2-\frac{H}{r_c}=\frac{8\pi{G}}{3}\rho_m~,
\end{equation}
where $r_c=(H_0(1-\Omega_m))^{-1}$ is the crossover scale. At early
times, $Hr_c\gg{1}$, the Friedmann equation of general relativity is
recovered, while in the future, $H\rightarrow{H_{\infty}}=1/r_c$,
the expansion is asymptotically de Sitter. Recently there have been
a lot of interests in the phenomenological studies relevant to the
DGP model in the literature \cite{Lue,DGPRev}.

In this note we investigate an interesting phenomenological model,
first introduced in Ref.\cite{mDGP}, which interpolates between the
pure $\Lambda$CDM model and the DGP model with an additional
parameter $\alpha$ and presents the tight constraints from the
current observations and future measurements. The paper is organized
as follows: In Sec. II we describe the general formalism of the
modified gravity model. Sec. III contains the current observations
we use, and Sec. IV includes our main global fitting results. In
Sec. V we present the forecasts from the future measurements, while
Sec. VI is dedicated the conclusions.


\section{General Formalism}

In this phenomenological model, assuming the flatness of our
universe, the Friedmann equation is modified as \cite{mDGP}:
\begin{equation}
H^2-\frac{H^\alpha}{r_c^{2-\alpha}}=\frac{8\pi{G}}{3}\rho_m~,\label{mDGPeq}
\end{equation}
where $r_c=H^{-1}_0/(1-\Omega_m)^{\alpha-2}$. Thus, we can
straightforwardly rewrite the above equation and obtain the
expansion rate as following:
\begin{equation}
E^2(z)\equiv\frac{H^2}{H^2_0}=\Omega_m(1+z)^3+\frac{\delta{H^2}}{H^2_0}~,\label{ez}
\end{equation}
where the last term denotes the modification of the Friedmann
equation of general relativity:
\begin{equation}
\frac{\delta{H^2}}{H^2_0}\equiv(1-\Omega_m)\frac{H^{\alpha}}{H^{\alpha}_0}=(1-\Omega_m)E^\alpha(z)~.\label{deltah}
\end{equation}
Furthermore, we can obtain the effective equation of state:
\begin{equation}
w_{\rm
eff}(z)\equiv-1+\frac{1}{3}\frac{d\ln{\delta{H^2}}}{d\ln(1+z)}
=-1+\frac{\alpha}{3}(1+z)\frac{E'(z)}{E(z)}~,\label{weff}
\end{equation}
where the prime denotes the derivative with respect to the redshift
$z$.

\begin{figure}[t]
\begin{center}
\includegraphics[scale=0.45]{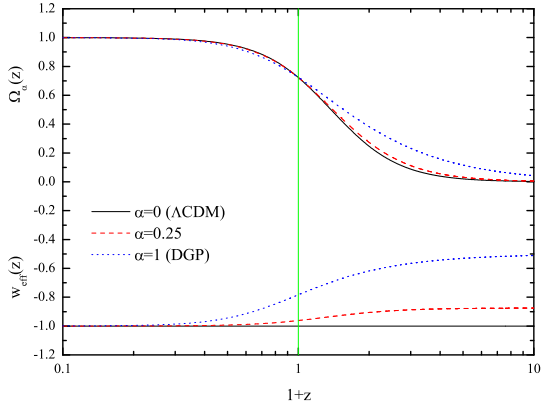}
\caption{The evolutions of effective energy density
$\Omega_\alpha(z)$ and effective equation of state $w_{\rm eff}(z)$
for different values of $\alpha$. The black solid lines are for
$\alpha=0$ ($\Lambda$CDM), the red dashed lines for $\alpha=0.25$
and the blue dotted lines for $\alpha=1$ (DGP). And the vertical
line denotes today ($z=0$).\label{fig1}}
\end{center}
\end{figure}

In Fig.\ref{fig1} we illustrate the evolutions of the effective
energy density $\Omega_\alpha(z)\equiv{1-\Omega_m(z)}$ and $w_{\rm
eff}(z)$ for different values of parameter $\alpha$. During the
matter dominated era, $E(z)$ varies as $(1+z)^{3/2}$, which
corresponds to the effective equation of state: $w_{\rm
eff}=-1+\alpha/2$. In the future $z\rightarrow0$, with the matter
density $\rho_m\propto{(1+z)^3}\rightarrow0$, we have $w_{\rm
eff}(z)\rightarrow-1$ and $\Omega_\alpha(z)\rightarrow1$.

Besides the modification of the Friedmann equation, the flat DGP
model also changes the growth function of density perturbation
$\delta(a)$. Under assumptions of a quasi-static regime and
sub-horizon scales, the correct evolution of perturbation was found
\cite{Lue,Koyama1}:
\begin{equation}
\ddot{\delta}+2H\dot{\delta}-4\pi{G}\left(1+\frac{1}{3\beta}\right)\rho_m\delta=0~,\label{groweq}
\end{equation}
where the dot denotes the derivative with respect to the cosmic time
$t$, and the $\beta$ factor is given by:
\begin{equation}
\beta=1-2r_cH\left(1+\frac{\dot{H}}{3H^2}\right)~.\label{beta1}
\end{equation}
However, this phenomenological model Eq.(\ref{mDGPeq}) is a
parametrization, so the situation is more complicated. One of the
possible methods was found by Ref.\cite{Koyama2}. In order to obtain
the growth function of density perturbation within a covariant
theory, the authors introduced a correction term and assumed the
structure of modified theory of gravity to determine this term.
Based on those assumptions, it was consequently found that the
$\beta$ factor was:
\begin{equation}
\beta=1-\frac{2}{\alpha}(Hr_c)^{2-\alpha}\left(1+(2-\alpha)\frac{\dot{H}}{3H^2}\right)~.\label{beta}
\end{equation}
In the following analysis, we will use Eq.(\ref{groweq}) and
Eq.(\ref{beta}) to calculate the growth of density perturbation.

\begin{figure}[t]
\begin{center}
\includegraphics[scale=0.45]{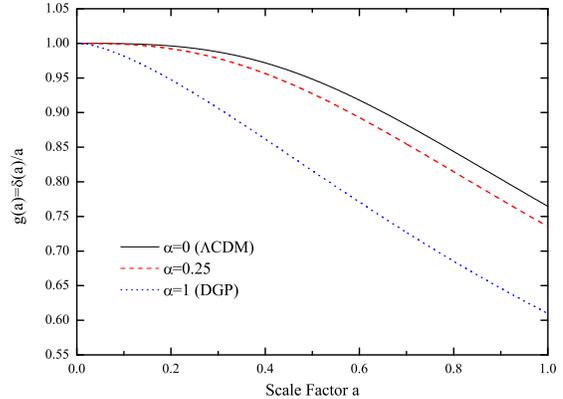}
\caption{The evolutions of linear growth $g(a)\equiv\delta(a)/a$ for
different values of $\alpha$. The black solid lines are for
$\alpha=0$ ($\Lambda$CDM), the red dashed lines for $\alpha=0.25$
and the blue dotted lines for $\alpha=1$ (DGP).\label{fig2}}
\end{center}
\end{figure}

Defined the normalized growth $g(a)\equiv\delta(a)/a$, the growth
function Eq.(\ref{groweq}) can be rewritten as:
\begin{eqnarray}
\frac{d^2g}{da^2}&+&\left(\frac{7}{2}-\frac{3}{2}\frac{w_{\rm
eff}(a)}{1+X(a)}\right)
\frac{dg}{ada}+\frac{3}{2}\left[1-\frac{w_{\rm eff}(a)}{1+X(a)}\right.\nonumber\\
&-&\left.\frac{X(a)}{1+X(a)}\left(1+\frac{1}{3\beta}\right)\right]\frac{g}{a^2}=0
\end{eqnarray}
where the variable $X(a)$ is the ratio of the matter density to the
effective energy density $X(a)=\Omega_m(a)/\Omega_\alpha(a)$. In
Fig.\ref{fig2} we plot the linear growth factor $g(a)$ as function
of scale factor $a$ for different values of $\alpha$. One can see
that the linear growth factor has been suppressed obviously as long
as $\alpha$ is larger than zero. Thus, in the literature the linear
growth has been widely used to study the modified gravity models,
especially the DGP model \cite{DGPGrowth}.

Furthermore, the growth factor can be parameterized as \cite{gamma}:
\begin{equation}
f\equiv\frac{d\ln\delta}{d\ln{a}}=\Omega_m^\gamma~,
\end{equation}
where $\gamma$ is the growth index. And then the growth function
becomes:
\begin{eqnarray}
\frac{df}{d\ln
a}&+&\left(\frac{1}{2}-\frac{3}{2}\frac{w_{\rm eff}(a)}{1+X(a)}\right)f+f^2\nonumber\\
&-&\frac{3}{2}\frac{X(a)}{1+X(a)}\left(1+\frac{1}
{3\beta}\right)=0~.
\end{eqnarray}
For the pure $\Lambda$CDM model, the theoretical value of $\gamma$
is $6/11\approx0.545$, while $\gamma=11/16=0.6875$ in the flat DGP
model \cite{gammavalue}.

In the framework of this phenomenological model, we can easily see
that the pure $\Lambda$CDM model and flat DGP model can be recovered
when $\alpha=0$ and $\alpha=1$, respectively. In order to be
consistent with the cosmological observations, the $\alpha$ term
should be very small in the early times, such as the Big Bang
Nucleosynthesis (BBN) era ($z\sim10^9$). This limit corresponds to
the upper bound: $\alpha<2$ \cite{mDGP}. On the other hand, when
$\alpha<0$, the effective equation of state will become smaller than
$-1$, which leads to the instability of linear growth of density
perturbation due to the appearance of $\beta$ term Eq.(\ref{beta}).

\begin{figure}[t]
\begin{center}
\includegraphics[scale=0.45]{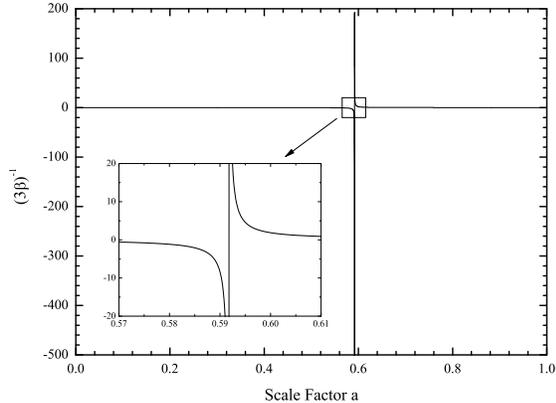}
\caption{The evolution of the $\beta$ term $(3\beta)^{-1}$ when
fixing $\alpha=-2$.\label{fig3}}
\end{center}
\end{figure}

At early times, such as the matter dominated era, we have
$E(z)\propto(1+z)^{3/2}$ and $\dot{H}/H^2\simeq-3/2$. Thus,
Eq.(\ref{beta}) becomes
\begin{equation}
\beta\simeq1-\frac{(\Omega_m(1+z)^3)^{1-\alpha/2}}{1-\Omega_m}\ll0~,\label{beta+}
\end{equation}
since $(1+z)^3\gg1$ at $2<z<1000$. By contrast, at late times the
matter energy density $\rho_m\propto{(1+z)^3}\rightarrow0$ and the
expansion is asymptotically de Sitter, $\dot{H}\rightarrow0$. And
then we have
\begin{equation}
\beta\simeq1-2/\alpha\Rightarrow\cases{\beta<0~,~~{\rm
for}~~0<\alpha<2\cr\beta>0~,~~{\rm for}~~\alpha<0\cr}~.\label{beta-}
\end{equation}
Based on Eq.(\ref{beta+}) and Eq.(\ref{beta-}), we can
straightforwardly see that as long as $\alpha<0$, during the
evolution of universe the value of $\beta$ should change the sign at
one pivot redshift $z_t$, which leads to $\beta|{}_{z_t}=0$ and
$(3\beta)^{-1}|_{z_t}\rightarrow\infty$. In Fig.\ref{fig3} we have
shown the evolution of $(3\beta)^{-1}$ when fixing $\alpha=-2$.
There is an obvious singularity at $a\sim0.595$. Therefore, based on
these discussions above, we use a tophat prior on $\alpha$ as
$0\leq\alpha<2$ in our calculations.


\section{Method and Data}

In our calculations we assume a flat space and use a uniform prior
on the present matter density fraction of the universe:
$0.1<\Omega_m<0.5$. Furthermore, we constrain the Hubble parameter
to be uniformly in $4\sigma$ Hubble Space Telescope (HST) region:
$0.4<h<1.0$. The resulting plots are produced with
CosmoloGUI\footnote{URL: http://www.sarahbridle.net/cosmologui/.}.

In this section we will list the cosmological observations used in
our calculations: CMB, BAO and SNIa measurements, as well as some
high-redshift observations, such as the GRB and LGF data. We have
taken the total likelihood to be the products of the separate
likelihoods ($\mathcal{L}_i$) of these cosmological probes. In other
words, defining $\chi^2_{L,i}=-2\log{\mathcal{L}_i}$, we get:
\begin{eqnarray}
\chi^2_{L,\rm total}&=&\chi^2_{L,\rm CMB}+\chi^2_{L,\rm
BAO}+\chi^2_{L,\rm SNIa}\nonumber\\
&+&\chi^2_{L,\rm GRB}+\chi^2_{L,\rm LGF}~.
\end{eqnarray}
If the likelihood function is Gaussian, $\chi^2_{L}$ coincides with
the usual definition of $\chi^2$ up to an additive constant
corresponding to the logarithm of the normalization factor of
$\mathcal{L}$.

\subsection{CMB Data}

CMB measurement is sensitive to the distance to the decoupling epoch
via the locations of peaks and troughs of the acoustic oscillations.
Here we use the ``WMAP distance information" obtained by the WMAP
group \cite{Komatsu:2008hk}, which includes the ``shift parameter"
$R$, the ``acoustic scale" $l_A$, and the photon decoupling epoch
$z_\ast$\footnote{In the revised version of WMAP5 paper
\cite{Komatsu:2008hk}, they also extend the baryon density $\Omega_b
h^2$ into the WMAP distance information. But our calculations are
not sensitive to $\Omega_b h^2$ and they also claim that this
extension does not affect the constraints. Thus, we fix $\Omega_b
h^2=0.022765$ to be the best fit value obtained by the WMAP group.}.
$R$ and $l_A$ correspond to the ratio of angular diameter distance
to the decoupling era over the Hubble horizon and the sound horizon
at decoupling, respectively, given by:
\begin{eqnarray}
R&=&\frac{\sqrt{\Omega_mH^2_0}}{c}\chi(z_\ast)~,\\
l_A&=&\frac{\pi\chi(z_\ast)}{\chi_s(z_\ast)}~,
\end{eqnarray}
where $\chi(z_\ast)$ and $\chi_s(z_\ast)$ denote the comoving
distance to $z_\ast$ and the comoving sound horizon at $z_\ast$,
respectively. The decoupling epoch $z_\ast$ is given by \cite{zast}:
\begin{equation}
z_\ast=1048[1+0.00124(\Omega_b h^2)^{-0.738}][1+g_1(\Omega_m
h^2)^{g_2}]~,
\end{equation}
where
\begin{equation}
g_1=\frac{0.0783(\Omega_b h^2)^{-0.238}}{1+39.5(\Omega_b
h^2)^{0.763}}~,~g_2=\frac{0.560}{1+21.1(\Omega_b h^2)^{1.81}}~.
\end{equation}
We calculate the likelihood of the WMAP distance information as
follows:
\begin{equation}
\chi^2=(x^{\rm th}_i-x^{\rm data}_i)(C^{-1})_{ij}(x^{\rm
th}_j-x^{\rm data}_j)~,
\end{equation}
where $x=(R,l_A,z_\ast)$ is the parameter vector and $(C^{-1})_{ij}$
is the inverse covariance matrix for the WMAP distance information
shown in Table I.

\begin{table}
Table I. Inverse covariance matrix for the WMAP distance information
$l_A$, $R$ and $z_{\ast}$. The maximum likelihood values are
$R=1.710$, $l_A=302.10$ and $z_\ast=1090.04$, respectively.
\begin{center}
\begin{tabular}{cccc}
  \hline
  \hline
    &~~~$l_A(z_{\ast})$~~~&~~$R(z_{\ast})$~~~&~~~~~$z_{\ast}$~~~\\
  \hline
  ~~$l_A(z_{\ast})$ & $1.800$ & $27.968$ & ~~$-1.103$\\
  ~~$R(z_{\ast})$ &  & $5667.577$ & ~~$-92.263$\\
  ~~$z_{\ast}$ &  & & ~~~$2.923$\\

  \hline
  \hline
\end{tabular}
\end{center}
\end{table}

\subsection{BAO Data}

The BAO information has been already detected in the current galaxy
redshift survey. The BAO can directly measure not only the angular
diameter distance, $D_A(z)$, but also the expansion rate of the
universe, $H(z)$. But current BAO data are not accurate enough for
extracting the information of $D_A(z)$ and $H(z)$ separately
\cite{Okumura:2007br}. Therefore, one can only determine the
following effective distance \cite{Eisenstein:2005su}:
\begin{equation}
D_v(z)\equiv\left[(1+z)^2D_A^2(z)\frac{cz}{H(z)}\right]^{1/3}~.
\end{equation}
In this note we use the gaussian priors on the distance ratios
$r_s(z_d)/D_v(z)$:
\begin{eqnarray}
r_s(z_d)/D_v(z=0.20)&=&0.1980\pm0.0058~,\nonumber\\
r_s(z_d)/D_v(z=0.35)&=&0.1094\pm0.0033~,
\end{eqnarray}
with a correlation coefficient of $0.39$, extracted from the SDSS
and 2dFGRS surveys \cite{BAO}, where $r_s(z_d)$ is the comoving
sound horizon size and $z_d$ is the drag epoch at which baryons were
released from photons given by \cite{zd}:
\begin{equation}
z_d=\frac{1291(\Omega_m h^2)^{0.251}}{1+0.659(\Omega_m
h^2)^{0.828}}[1+b_1(\Omega_b h^2)^{b_2}]~,
\end{equation}
where
\begin{eqnarray}
b_1&=&0.313(\Omega_m h^2)^{-0.419}[1+0.607(\Omega_m
h^2)^{0.674}]~,\nonumber\\
b_2&=&0.238(\Omega_m h^2)^{0.223}~.
\end{eqnarray}

\subsection{SNIa Data}

The SNIa data give the luminosity distance as a function of redshift
\begin{equation}
d_L=(1+z)\int^{z}_0\frac{cdz'}{H(z')}~.
\end{equation}
The supernovae data we use in this paper are the recently released
Union compilation (307 sample) from the Supernova Cosmology project
\cite{Kowalski:2008ez}, which include the recent samples of SNIa
from the SNLS and ESSENCE survey, as well as some older data sets,
and span the redshift range $0\lesssim{z}\lesssim1.55$. In the
calculation of the likelihood from SNIa we have marginalized over
the nuisance parameter, the absolute magnitude $M$, as done in
Ref.\cite{SNMethod}:
\begin{equation}
\bar{\chi}^2=A-\frac{B^2}{C}+\ln\left(\frac{C}{2\pi}\right)~,
\end{equation}
where
\begin{eqnarray}
A&=&\sum_i\frac{(\mu^{\rm data}-\mu^{\rm th})^2}{\sigma^2_i}~,\nonumber\\
B&=&\sum_i\frac{\mu^{\rm data}-\mu^{\rm th}}{\sigma^2_i}~,\nonumber\\
C&=&\sum_i\frac{1}{\sigma^2_i}~.
\end{eqnarray}

\subsection{GRB Data}

GRBs can potentially be used to measure the luminosity distance out
to higher redshift than SNIa. Recently, several empirical
correlations between GRB observables were reported, and these
findings have triggered intensive studies on the possibility of
using GRBs as cosmological ``standard'' candles. However, due to the
lack of low-redshift long GRBs data to calibrate these relations, in
a cosmology-independent way, the parameters of the reported
correlations are given, assuming an input cosmology, and obviously
they depend on the same cosmological parameters that we would like
to constrain. Thus, applying such relations to constrain
cosmological parameters leads to biased results. In
Ref.\cite{xiagrb} the circular problem is naturally eliminated by
marginalizing over the free parameters involved in the correlations;
in addition, some results show that these correlations do not change
significantly for a wide range of cosmological parameters
\cite{Firmani}. Therefore, in this paper we use the 69 GRBs sample
over a redshift range from $z=0.17-6.60$ published in
Ref.\cite{Schaefer:2006pa} but we keep in mind the issues related to
the ``circular problem" that are more extensively discussed in
Ref.\cite{xiagrb}.

\subsection{LGF Data}

As we point out above, the linear growth factor will be suppressed
in the modified gravity model. It will be helpful using the
measurements of linear growth factor to constrain the modified
gravity models. Therefore, in Table II we list linear growth factors
data we use in our analysis: the linear growth rate
$f\equiv\Omega^\gamma_m$ from galaxy power spectrum at low redshifts
\cite{Hawkins:2002sg,Tegmark:2006az,Ross:2006me,Guzzo,daAngela:2006mf}
and lyman-$\alpha$ growth factor measurement obtained with the
lyman-$\alpha$ power spectrum at $z=3$ \cite{McDonald:2004xn}. It is
worth noting that the data points in Table II are obtained with
assuming the $\Lambda$CDM model, thus, one should use these data
very carefully, especially for the points obtained from
Refs.\cite{Tegmark:2006az,Ross:2006me,daAngela:2006mf}. The
corresponding $\chi^2$ is simply given by:
\begin{equation}
\chi^2=\sum_i\frac{(f^{\rm th}_i-f^{\rm data}_i)^2}{\sigma^2_i}~.
\end{equation}

\begin{table}
TABLE II. The currently available data for linear growth rates $f$
we use in our analysis.
\begin{center}

\begin{tabular}{cccc}







\hline\hline

~~~~$z$~~~~ & ~~~~$f$~~~~ &~~~~$\sigma$~~~~& ~~Ref.~~~\\

\hline

0.15&0.51&0.11&\cite{Hawkins:2002sg}\\
0.35&0.70&0.18&\cite{Tegmark:2006az}\\
0.55&0.75&0.18&\cite{Ross:2006me}\\
0.77&0.91&0.36&\cite{Guzzo}\\
1.40&0.90&0.24&\cite{daAngela:2006mf}\\
3.00&1.46&0.29&\cite{McDonald:2004xn}\\

\hline\hline
\end{tabular}
\end{center}
\end{table}


\section{Numerical Results}

\begin{table}
TABLE III. Constraints on the parameters $\alpha$, $\Omega_m$ and
$\gamma$. Here we have shown the mean values and errors from the
current observations and the standard derivations from the future
measurements.

\begin{center}

\begin{tabular}{cccc}

\hline\hline

&$\alpha$&$\Omega_m$&$\gamma$\\

\hline

CMB+BAO+SN & $0.263\pm0.175$ & $0.276\pm0.018$ & $-$\\
All Real Data&$0.254\pm0.153$&$0.277\pm0.017$&$0.570\pm0.205$\\
Future&0.07&0.005&0.050\\

\hline\hline
\end{tabular}
\end{center}
\end{table}

\begin{figure}[t]
\begin{center}
\includegraphics[scale=0.23]{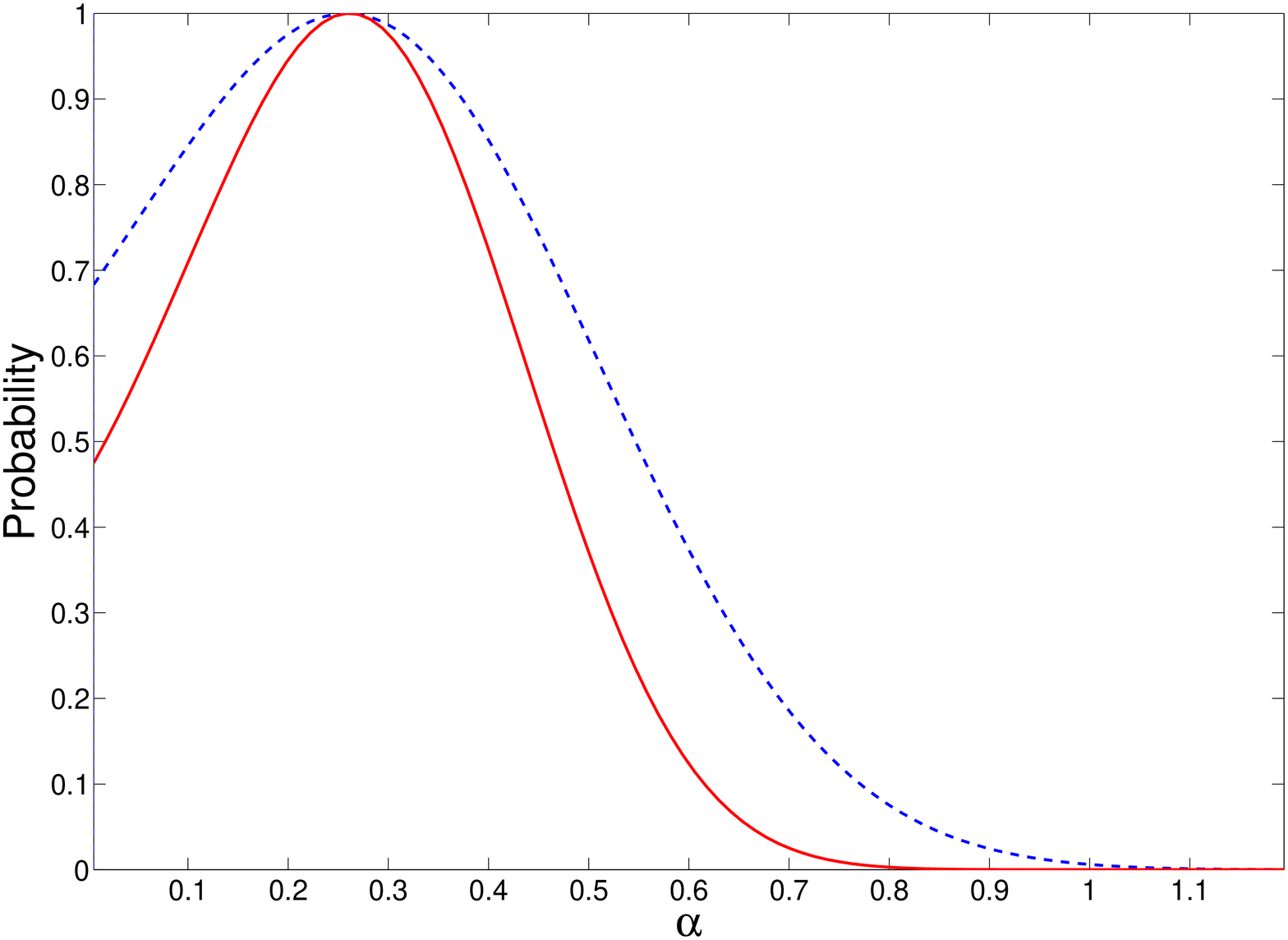}
\caption{One dimensional constraint on the parameter $\alpha$ from
the different current data combinations: CMB+BAO+SN (blue dashed
lines) and all real data (red solid lines). \label{fig4}}
\end{center}
\end{figure}

In this section we present our main results of constraints on this
phenomenological model from the current observational data, as shown
in Table III.

In Fig.\ref{fig4} we illustrate the posterior distribution of
$\alpha$ from the current data. Firstly, we neglect the high
redshift probes. The result shows that the current observations
yield a strong constraint on the parameter:
\begin{equation}
\alpha=0.263\pm0.175~(1\sigma)~.\label{result1}
\end{equation}
One can see that the pure $\Lambda$CDM model ($\alpha=0$) still fits
data very well at $2\sigma$ uncertainty, which is consistent with
the current status of global fitting results
\cite{Komatsu:2008hk,xiaglobal}. And the $95\%$ upper limit is
$\alpha<0.686$, which implies that there is a significant tension
between the flat DGP model ($\alpha=1$) and the current
observations, which is consistent with other works (e.g.
Ref.\cite{ariel,otherDGP}). However, unlike other works
\cite{ariel}, in our analysis we use the ``WMAP distance
information" which includes the ``shift parameter" $R$, the
``acoustic scale" $l_A$, and the photon decoupling epoch $z_\ast$,
instead of $R$ only, to constrain this phenomenological model.
Recently, many results show that the ``WMAP distance information"
can give the similar constraints, when compared with the results
from the full CMB power spectrum \cite{compare}. By contrast, $R$
could not be an accurate substitute for the full CMB data and may in
principle give some misleading results \cite{shift}.

And then, we include some high redshift probes, such as GRB and LGF
data sets. From Table III and Fig.\ref{fig4}, we can find that the
constraint on $\alpha$ becomes slightly tighter:
\begin{equation}
\alpha=0.254\pm0.153~(1\sigma)~,\label{result2}
\end{equation}
and $\alpha<0.541$ at $2\sigma$ confidence level. As we have
mentioned before, the effective equation of state of this
phenomenological model will depart from the cosmological constant
boundary at high redshifts. Therefore, these high redshift
observations are helpful to improve the constraints on this
phenomenological model.

These results (Eqs.(\ref{result1}-\ref{result2})) are not
surprising. From Fig.\ref{fig1} we find that the effective equation
of state of the flat DGP model, $w_{\rm
eff}\approx-1+\alpha/2=-0.5$, will depart from the cosmological
constant $w=-1$ at high-redshift universe significantly. But the
current constraint on $w$ is closed to $w=-1$
\cite{Komatsu:2008hk,xiaglobal}, so we require the small value of
$\alpha$ to match the current observations. There is a small
difference that our result slightly favors a non-zero value of
$\alpha$, but not significantly, which needs more accurate
measurements to verify it further.

\begin{figure}[t]
\begin{center}
\includegraphics[scale=0.2]{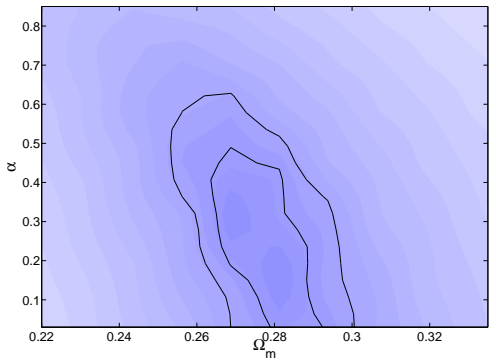}
\caption{$68\%$ and $95\%$ constraints in the ($\Omega_m$,$\alpha$)
plane from the current observations.\label{fig5}}
\end{center}
\end{figure}

In Fig.\ref{fig5} we plot the two dimensional constraint in the
($\Omega_m$,$\alpha$) panel. $\Omega_m$ and $\alpha$ are strongly
anti-correlated. The reason of this degeneracy is that the
constraint mainly comes from the luminosity and angular diameter
distance information. From Eq.(\ref{ez}) and Eq.(\ref{deltah}) we
can see that when $\alpha$ is increased, the contribution of last
$\alpha$ term to the expansion rate will become large, due to the
positive $E(z)$. Consequently, $\Omega_m$ must be decreased
correspondingly in order to produce the same expansion rate. When
combining those current observational data, the matter energy
density has been constrained very stringent:
$\Omega_m=0.277\pm0.017~(1\sigma)$, which is also consistent with
the current status of global fitting results
\cite{Komatsu:2008hk,xiaglobal}. Naturally, the constraint on
$\alpha$ will also be improved, because of the tight constraint on
the matter energy density.

Furthermore, we also investigate the limit on the growth index
$\gamma$ and obtain $\gamma=0.570\pm0.205$ at $68\%$ confidence
level. Obviously, the growth index of the pure $\Lambda$CDM
$\gamma=6/11\approx0.545$ is consistent with this result. However,
the theoretical value of growth index in the flat DGP model,
$\gamma=11/16=0.6875$, is disfavored.


\section{Future Constraints}

Since the present data clearly do not give very stringent constraint
on the parameter $\alpha$, it is worthwhile to discuss whether
future data could determine $\alpha$ conclusively. For that purpose
we have performed an analysis and chosen the fiducial model as the
mean values of Table III obtained from the current constraints.

The projected satellite SNAP (Supernova / Acceleration Probe) would
be a space based telescope with a one square degree field of view
with $10^9$ pixels. It aims to increase the discovery rate for SNIa
to about $2000$ per year in the redshift range $0.2<z<1.7$. In this
paper we simulate about $2000$ SNIa according to the forecast
distribution of the SNAP \cite{Kim:2003mq}. For the error, we follow
the Ref.\cite{Kim:2003mq} which takes the magnitude dispersion
$0.15$ and the systematic error $\sigma_{\rm sys}=0.02\times z/1.7$.
The whole error for each data is given by:
\begin{equation}
\sigma_{\rm mag}(z_i)=\sqrt{\sigma^2_{\rm
sys}(z_i)+\frac{0.15^2}{n_i}}~,\label{snap}
\end{equation}
where $n_i$ is the number of supernovae of the $i'$th redshift bin.
Furthermore, we add as an external data set a mock dataset of 400
GRBs, in the redshift range $0 < z < 6.4$ with an intrinsic
dispersion in the distance modulus of $\sigma_{\mu}=0.16$ and with a
redshift distribution very similar to that of Figure 1 of
Ref.\cite{Hooper:2005xx}.

\begin{figure}[t]
\begin{center}
\includegraphics[scale=0.2]{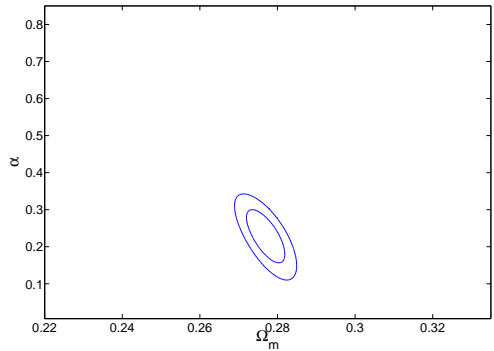}
\caption{$68\%$ and $95\%$ constraints in the ($\Omega_m$,$\alpha$)
plane from the future measurements.\label{fig6}}
\end{center}
\end{figure}

For the linear growth factors data, we simulate the mock data from
the fiducial model with the error bars reduced by a factor of two.
This is probably reasonable given the larger amounts of galaxy power
spectrum and lyman-$\alpha$ forest power spectrum data that will
become available soon as long with a better control of systematic
errors in the next generated large scale structure survey. In
addition we also assume a Gaussian prior on the matter energy
density $\Omega_m$ as $\sigma=0.007$, which is close to future
Planck constraints \cite{Planck}.

From Table III it is clear that the future measurements with higher
precision could improve the constraints dramatically. The standard
derivation of $\alpha$ is reduced by a factor two. Assuming the mean
value remains unchanged in the future, the non-zero value of
$\alpha$ will be confirmed around $3\sigma$ confidence level by the
future measurements. In addition we also illustrate the two
dimensional contour of parameters $\Omega_m$ and $\alpha$ in
Fig.\ref{fig6}. Comparing with the contour in Fig.\ref{fig5}, the
allowed parameter region has been shrunk significantly. The future
measurements could have enough ability to distinguish between the
modified gravity model and the pure $\Lambda$CDM model.


\section{Conclusions}

As an alternative approach to generate the late-time acceleration of
the expansion of our universe, models of modifications of gravity
have attracted a lot of interests in the phenomenological studies
recently. In this note we investigate an interesting
phenomenological model which interpolates between the pure
$\Lambda$CDM model and the flat DGP braneworld model with an
additional parameter $\alpha$.

Firstly, we find that when $\alpha$ is less than zero, the growth
function of density perturbation $\delta(a)$ will appear an apparent
singularity. This is because the variable $\beta$ will change the
sign during the evolution of our universe. And then the $\beta$ term
caused by the modified gravity model will be divergent at some
redshift $z_t$.

From the current CMB, BAO and SNIa data, we obtain a tight
constraint on the parameter $\alpha=0.263\pm0.175~(1\sigma)$, which
implies that the flat DGP model ($\alpha=1$) is incompatible with
the current observations, while the pure $\Lambda$CDM model still
fits the data very well. When adding the high-redshift GRB and LGF
data, the constraint is more stringent
$\alpha=0.254\pm0.153~(1\sigma)$, which means that these high
redshift observations are helpful to improve the constraints on this
phenomenological model.

Finally, we simulate the future measurements with higher precisions
to limit this phenomenological model. And we find that these
accurate probes will be helpful to improve the constraints on the
parameters of the model and could distinguish between the modified
gravity model and the pure $\Lambda$CDM model.





\end{document}